\documentclass[12pt,pra,aps]{revtex4}
\usepackage{graphicx}
\usepackage{rotating}
\usepackage{array}
\usepackage{multirow}
\usepackage{epstopdf}
\usepackage{amsmath}
\usepackage{amsfonts}
\usepackage{setspace}
\usepackage{braket}
\usepackage{color}
\usepackage{nameref,zref-xr} 
\usepackage{hyperref}

\begin{document}
\title{Is solid copper oxalate a spin chain or a mixture of entangled spin pairs?}

\author{Pavel Pokhilko$^a$, Dmitry S. Bezrukov$^{b,c}$, and Anna I. Krylov$^a$\\
{\small $^a$ Department of Chemistry, University of Southern California, 
  Los Angeles,  California 90089-0482}\\
{\small $^b$ Department of Chemistry, M.V. Lomonosov Moscow State University, Moscow 119991, Russia}\\
{\small $^c$ Skolkovo Institute of Science and Technology, Skolkovo Innovation Center, Nobel str. 3, Moscow 143026, Russia}\\
}

\begin{abstract}
  Macroscopic assemblies of interacting spins give rise to a broad spectrum of behaviors
  determined by the spatial arrangement of the magnetic sites and the electronic interactions between them.
  Compounds of copper (II), in which each copper carries spin $\frac{1}{2}$,
  exhibit a vast 
  variety of physical properties. For antiferromagnetically coupled spin sites, there are 
  two limiting scenarios: spin chains in which the spins can exhibit a long-range order
  or a mixture of dimers in which the spins within each pair are entangled but
  do not communicate with the spins from other dimers. In principle, 
  the two types can be distinguished on the  basis of experimental observations and
  modeling using empirically parameterized effective Hamiltonians,
  but in practice, ambiguity may persist for decades, as is the case for copper oxalate. 
  Here we use high-level \emph{ab initio} calculations to establish the validity of the nearest-site Heisenberg model and to predict the interaction strength between the magnetic sites. 
  The computed magnetic susceptibility provides an unambiguous interpretation of 
  magnetic experiments performed throughout half a century, clearly supporting the infinite spin-chain
  behavior of solid copper oxalate. 
\end{abstract}
\maketitle

\section{Introduction}

\begin{figure}[h!]
\includegraphics[width=11cm]{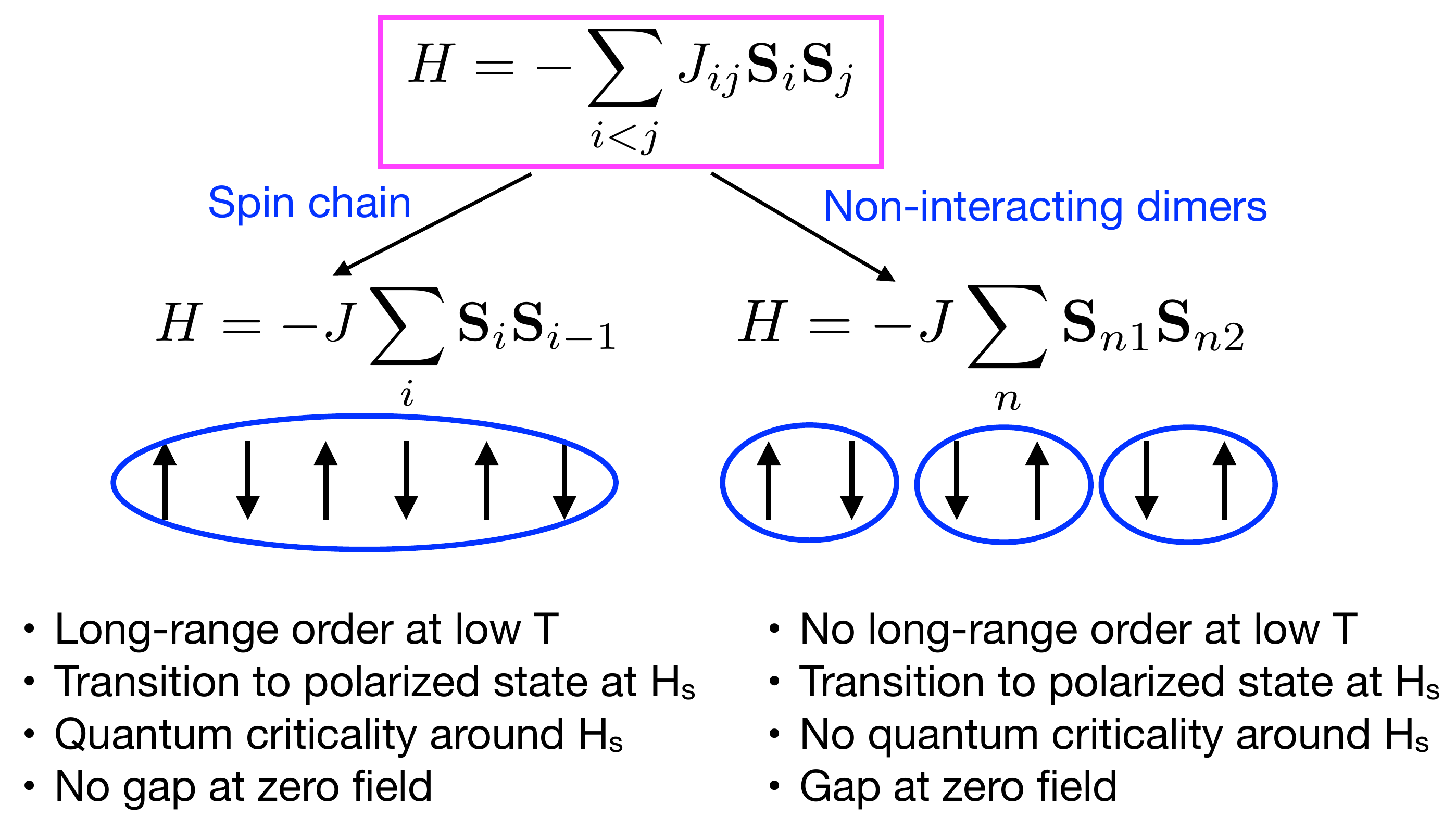} 
\centering
\caption{An assembly of interacting spins is described by the Heisenberg Hamiltonian. 
  Depending on the structure and strength of interactions, such system can exhibit vastly different physics.  For one-dimensional systems, the  two limiting cases are  spin chains
  in which the spins can exhibit a long-range order
  or a mixture of dimers in which the spins within each pair are entangled but
  do not communicate with the spins from other dimers.
\protect\label{fig:intro}}
\end{figure}

Strongly correlated materials exhibit unusual properties, which are exploited 
in emerging applications such as quantum information science and
spintronics\cite{Wunderlich2016:review:antif_spintronics}. Physical
properties depend on the spatial arrangement  and the strengths of electronic
couplings between the magnetic sites, as illustrated in  Fig. \ref{fig:intro}. For example,
chains of antiferromagnetically coupled spins exhibit long-range spin correlations and a zero spectral gap
at zero magnetic field. In contrast, an assembly of weakly coupled spin pairs has no long-range
order and has a gap at zero field. Both cases give rise to a bell-shape magnetic susceptibility 
curve $\chi(T)$. 
In both types of systems, the equilibrium state can be converted to a fully spin-polarized state above saturation field,
roughly equal to the exchange coupling. 
The differences between the two cases
can be seen at low temperatures, e.g., spin chains have finite $\chi(T)$, in contrast to dimers.
The spectral gap and long-range order can also be probed by neutron scattering experiments, e.g.,
dynamic structural factors for the spin-chains at zero field start at zero energy and show
characteristic sinusoidal shape, following the spinon dispersion relation $\epsilon(k) = \frac{\pi}{2}|J| \sin k$,
as was illustrated for
copper sulfate\cite{Mourigal:CopperSulfate:spinon:2013};
whereas the dimers are expected to show flat bands at finite energy. 
In addition to their technological relevance\cite{Wunderlich2016:review:antif_spintronics}, 
spin chains are also fundamentally interesting---for example, they give rise to quantum
criticality around the saturation field\cite{Schofield:QuanumCriticality:05,Lorenz:2017:Cu:spin_chain}.

Copper (II) salts can behave both as antiferromagnetic spin chains\cite{Lorenz:2017:Cu:spin_chain,Mourigal:CopperSulfate:spinon:2013} and
as spin-paired dimers\cite{Bowers:mag_sus:fitting:1952,CuAc:Neese:2011}.
When experimental data is limited to finite-temperature  magnetic susceptibility,
as in the case of copper oxalate, one cannot confidently distinguish between the two regimes
without first-principle calculations. Here we report the first fully {\em ab initio} determination
of the effective Hamiltonian and macroscopic magnetic susceptibility for copper oxalate. The results 
provide unambiguous interpretation of  magnetic experiments performed throughout half a century
and clearly support the infinite spin chain behavior of the solid copper oxalate.\\

As shown in Fig. \ref{fig:intro}, an infinite chain of spins can be treated with the XXX Heisenberg
Hamiltonian 
\begin{gather}
  H = -J\sum_i \mathbf{S}_i \mathbf{S}_{i+1},
\protect\label{eq:XXX}  
\end{gather}
where the $\mathbf{S}_i$ are local spins and $J$ is the effective exchange constant. 
If the effective exchange constant is negative, 
the system adopts the antiferromagnetic singlet ground state with 
opposite spin orientation of the adjacent magnetic centers. 
The case of $S=1/2$ is one of the  few known quantum integrable models.
Its exact solution is given by the Bethe ansatz\cite{Bethe:ansatz:1931}, 
which facilitated the development of solid-state physics 
and mathematics of integrable systems. 
The thermodynamic properties of this model are well-known and have been used 
to explain experimental observables of real materials, 
containing, for instance, copper\cite{Lorenz:2017:Cu:spin_chain,Uchida:1996:Cu:spin_chain}, 
vanadium\cite{Jacobson:1987:V:spin_chain}, and magnesium\cite{Morosin:1973:Mn:spin_chain} magnetic centers. 

Despite its simplicity, the magnetic structure of copper oxalate has not been settled. 
On the basis of susceptibility measurements, EPR, and EXAFS, 
both a dimer\cite{Cu:oxalate:1966,Dubicki:Cu:chains:1966,Grivel:CuC2O4:cryst} and
an infinite spin chain Heisenberg models\cite{Soos:CuC2O4:EPR:1976,Goulon:Cu:EXAFS:1979} have been
proposed and used to fit the experimental data. Even after the determination of the crystal structure
in 2014,  the choice of magnetic model remained open\cite{Grivel:CuC2O4:cryst}.

Even when the model is known, the  first-principle determination of the effective exchange
constant is not trivial. The strongly correlated nature of magnetic systems
combined with their large sizes, pose a formidable challenge for quantum chemistry.
Here we use state-of-the-art wave-function methods based on equation-of-motion coupled-cluster (EOM-CC)
theory\cite{Krylov:EOMRev:07}. 
In contrast to popular broken-symmetry
density functional theory, our approach has no empirical or system-dependent parameters and does not
rely on unphysical spin-scrambled solutions. 
We begin with the full-electron treatment of model systems and
use the resulting wave functions to build
effective Hamiltonians\cite{Marlieu:MagnetRev:2014,Pokhilko:EffH:2020}, which afford a coarse-grained
description of the electronic structure and allow extrapolation to infinite systems. 
This is the first application of the spin-flip EOM-CC method to describe
a periodic system, opening a new route in the treatment of periodic strongly
correlated systems.
Our theoretical results provide unambiguous interpretation of the magnetic measurements\cite{Cu:oxalate:1966,Dubicki:Cu:chains:1966,Soos:CuC2O4:EPR:1976,Goulon:Cu:EXAFS:1979,Grivel:CuC2O4:cryst} 
of copper oxalate performed throughout several decades.

\section{Results and discussion}
\begin{figure}[!h]
  \includegraphics[width=11cm]{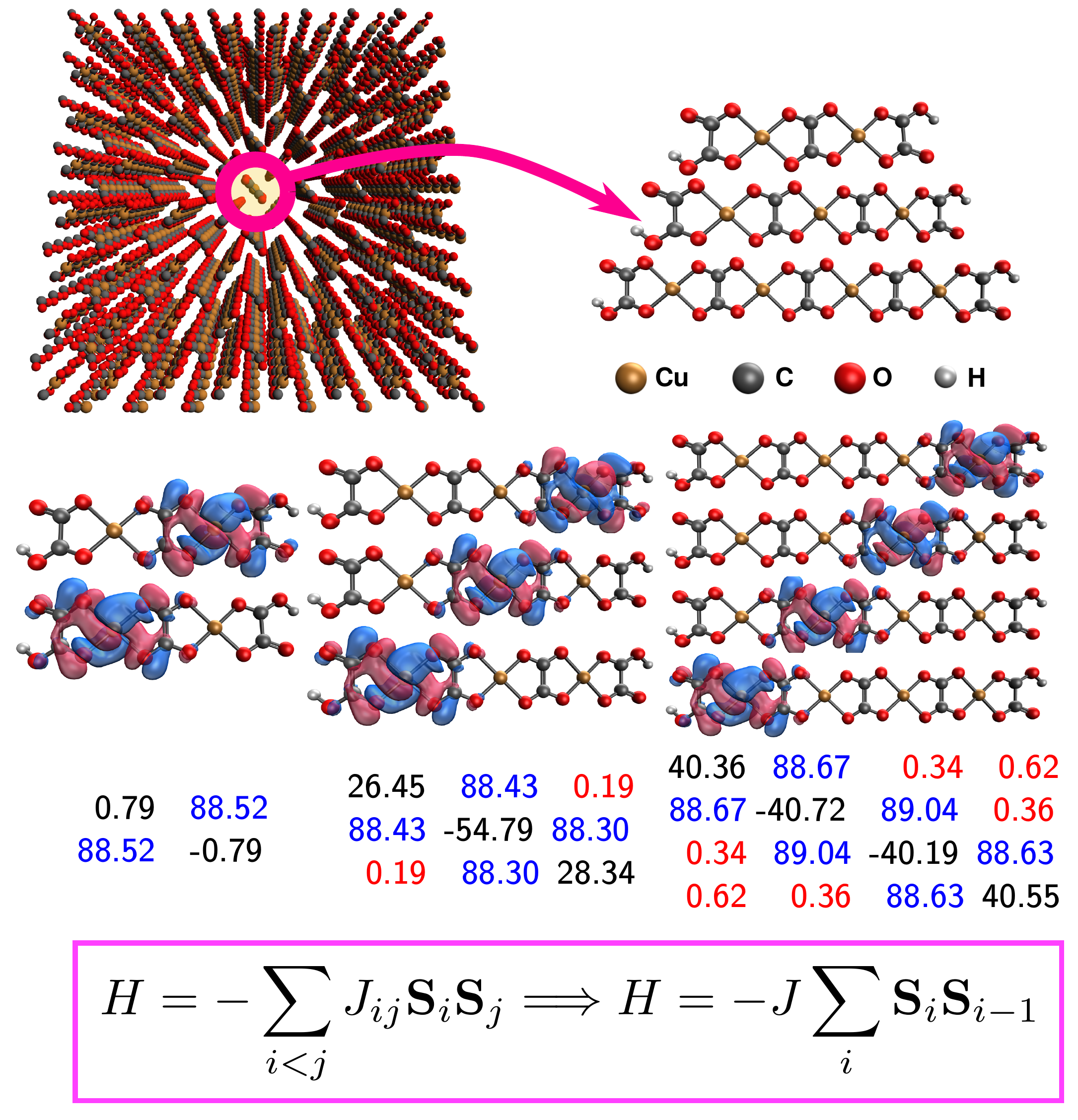} 
\centering
\caption{Top, left: Crystal structure of copper oxalate (CuC$_2$O$_4$). 
         Top, right: fragments with 2, 3, and 4 copper atoms from the crystal. 
         Middle: Localized orbitals defining the open-shell model spaces.
         Bottom: Des Cloizeaux effective Hamiltonians built from the EOM-SF wave functions 
         in the basis of open-shell determinants. 
         The energies are shifted to produce a zero trace. 
         The numbers in \textcolor{blue}{blue} show the couplings between nearest neighbors. 
         The numbers in \textcolor{red}{red} show the couplings between distant centers.
  \protect\label{fig:all_geometries}}
\end{figure}

The crystal structure\cite{Grivel:CuC2O4:cryst}
of copper oxalate, shown in Figure~\ref{fig:all_geometries}, reveals a regular pattern of 
copper oxalate chains with a bidentate orientation of the oxalate ligands. 
We begin with EOM-SF-CCSD/cc-pVDZ calculations 
for the  fragments of increasing length (containing 2, 3, and 4 copper centers, as  
shown in Figure~\ref{fig:all_geometries}, and capped with hydrogen atoms).
We use the resulting energies and wave functions to build
effective 
Hamiltonians\cite{Marlieu:MagnetRev:2014,Pokhilko:EffH:2020}, which afford a coarse-grained
description of the electronic structure.

The effective Heisenberg Hamiltonians 
for fragments  built from these highly accurate EOM-SF-CCSD calculations  
are shown in the middle panel of Figure~\ref{fig:all_geometries}.
Bloch's and des Cloizeaux's versions of the effective Hamiltonians are consistent 
with each other and yield nearly the same values of effective exchange constants $J$, 
as shown in Table~S2 %\ref{SI-tbl:J} 
in the SI. 
This procedure provides a rigorous mapping between the full-electron description and 
effective interactions between the magnetic centers.
In particular, the resulting effective exchange constant includes not only the contribution
from the exchange integral,  but also screening and correlation effects, folded in by means of 
the Bloch wave operator. 
We then use the Heisenberg Hamiltonian
to obtain the states that are not reachable by a single spin-flip, as was pioneered by Mayhall 
and Head-Gordon for single-molecule magnets\cite{Mayhall:1SF:2015}.

Note that in these calculations we constructed the full Heisenberg
Hamiltonian (as shown in the pink box in Fig. \ref{fig:intro}),  i.e., without
assuming nearest-neighbor approximation or constraining all
neighboring $J$ to be the same.
The analysis of these {\em ab initio} constructed  effective Hamiltonians
clearly reveals antiferromagnetic spin interactions.
The off-diagonal elements decay rapidly with the distance between the copper centers, such 
that only the nearest-neighbor spin couplings are significant. 
Non-nearest-neighbor couplings are about 1~cm$^{-1}$ or less, 
which is comparable with the thresholds used in the calculations. To further quantify the impact of neglecting
the non-nearest neighbors contributions, we carried out two additional numeric tests.
First, we computed eigenvalues of the Heisenberg Hamiltonian in which we zeroed the contributions from
non-nearest neighbors and compared them with the eigenvalues of the full Hamiltonian; the results are shown
in Table S3 of the SI. As one can see, the individual eigenvalues change by less than 0.5 cm$^{-1}$ and
the effect on the spectral gap is of the same magnitude---for example, the difference between the highest and the lowest eigenstate in the 4-copper Hamiltonian changes by 0.44 cm$^{-1}$ due to the zeroing out non-nearest neighbor
elements, which is indeed small compared  to the value of the gap (301.07 cm$^{-1}$). 
Second, we tri-diagonalized the full effective Hamiltonian for the 3 and 4 copper centers, so that the effect of
non-nearest neighbor contributions is folded into the remaining matrix elements. The results, which are shown in Fig. S2 in the SI, show that this operation changes the value of the key matrix element (between the two middle coppers) by
only 0.63 cm$^{-1}$. 
Therefore, we conclude that the nearest-neighbor approximation is indeed fully justified for this system, such that
the full Heisenberg Hamiltonian for the fragments can be replaced 
by the XXX Heisenberg Hamiltonian for the infinite spin chain, Eq. (\ref{eq:XXX}), 
which we use to extrapolate the Hamiltonian to the limit of the infinite number of copper centers. 
The effective exchange constants computed for the fragments of increasing sizes (collected in
Table~S2) %\ref{SI-tbl:J})
 show little variation and converge rapidly  with respect to the model
system size. 
Hence, for calculations of thermodynamic properties we use the $|J|$ constant from the middle of the 4-copper fragment (-178.1~cm$^{-1}$).

To account for weak correlations beyond EOM-CCSD,
we calculated a perturbative triples correction using the (fT) model for a model dimer.
Surprisingly, inclusion of triple excitations leads
to a more than 20\% increase of $|J|$ (42.7~cm$^{-1}$), 
illustrating that weak correlations are important for quantitative results. 
The direction of the change can be explained by the Pauli repulsion principle.
Pauli repulsion, which is already built-in in the
wave-function at the mean-field level by virtue of using Slater determinants,
allows the electrons in the triplet state to avoid each other.
The resulting Pauli hole covers the errors due to an incomplete description of the Coulomb hole,
which requires electron correlation. Consequently, the correlation effects are always
smaller for triplet states\cite{sfpaper} and improving correlation treatment results in
stabilizing singlets relative to triplets.
Previous calculations of dicopper single-molecule magnets suggested that the cc-pVDZ basis
is nearly sufficient\cite{Orms:magnets:17,Pavel:OSFNO:2019}.
To account for a basis-set effect beyond cc-pVDZ, we performed EOM-SF-CCSD calculations for the dimer with the cc-pVTZ basis set. These calculations show that this improvement of the basis-set reduces the $|J|$ value by 8.5~cm$^{-1}$. 

To account for spin--orbit interaction, which can split and shift energy levels, we computed spin--orbit
couplings (for the model dimer) using the Breit--Pauli Hamiltonian within spin--orbit mean-field approximation and
non-relativistic EOM-CC states. 
In these calculations we used the EOM-DIP-CCSD ansatz,
which can describe not only the lowest singlet and
triplet states (i.e., the Heisenberg states), but the entire manifold of the low-lying
states derived from different occupations
of coppers' $d$-orbitals. Although these states are relatively far in energy from the lowest
singlet/triplet pair (about 1.6 eV above the lowest singlet), they are expected to be important for
spin--orbit coupling by
virtue of El-Sayed's rule\cite{ElSayedRule-orig,ElSayedRule,Pavel:SOCNTOs:2019},
which says that states with different orbital occupancies give rise to 
large spin--orbit interactions.

The EOM-DIP-CCSD ansatz, which can handle strong correlation well but underestimates weak correlation and orbital relaxation, overestimates the energy gap relative to EOM-SF-CCSD
($J=-388.6$~cm$^{-1}$ for the dimer). To correct for this limitation, we combined the EOM-SF-CCSD energy gap 
with EOM-DIP-CCSD spin--orbit couplings to evaluate the effective $|J|$.
Table~S4 %\ref{SI-tbl:soc}
 illustrates 
the convergence of the $J$ constant with respect to the the lowest singlet number of interacting states.
We observe that both the lowest singlet and the lowest triplet states interact strongly through the
spin--orbit operator with higher excited states. 
This interaction shifts both states, largely canceling the effect of the spin--orbit coupling
on the singlet-triplet gap.
Nevertheless, 
the singlet interacts more strongly with the excited states than the triplet does, which   
results in a substantial  (13 \%)
increase of antiferromagnetic $|J|$ by $32$~cm$^{-1}$.
We note that these results indicate that neglect of the spin--orbit interaction
in previous studies of di-copper single-molecule magnets might be responsible for the reported
systematic
discrepancies between the theoretical and computed $J$-values\cite{Orms:magnets:17}.

Table \ref{tbl:J} summarizes the various contributions to the effective $J$ and gives our best estimate of
$J=-244$~cm$^{-1}=30.25$~meV. Comparing to other copper spin chains,  this value is 33 times higher
than that of copper pyrazine\cite{Lorenz:2017:Cu:spin_chain} 
and 118 times higher than of copper sulfate\cite{Mourigal:CopperSulfate:spinon:2013}, implying that a much stronger
magnetic field (on the order of 400-500 T) would be required to bring this system to the
fully polarized spin state and into the quantum criticality regime.

\begin{table}[tbh!]
  \caption{Contributions to the effective exchange constant $J$, cm$^{-1}$.
\protect\label{tbl:J}} 
\begin{tabular}{p{1.5cm}l}
$-178.1$    & Converged effective Hamiltonian \\
$-42.7$     & Perturbative triples\\
$-32$       & Spin--orbit shift\\
$+8.5$      & Basis-set effects\\
\cline{1-1}
$-244$      & Total
\end{tabular}
\end{table}

To compare the computed effective exchange constants with the experimental observable,
one needs to compute macroscopic magnetic susceptibility $\chi(T)$. 
Even with the solvable Bethe ansatz, such calculations require additional approximations.  
Here we consider two approaches: analytic classical and numeric quantum treatment of the temperature dependence of
$\chi(T)$. 
In the classical treatment, the quantum Heisenberg Hamiltonian is mapped onto the
classical Heisenberg Hamiltonian\cite{Fisher:class:spin_chain:1964}. 
The connection between the quantum and classical parameters in the case of $S=1/2$ is given by
\begin{gather}
J^{quant}/2 = J^{cl}, \\
g^{quant} = g^{cl}.
\end{gather}
The analytic solution of the $\chi(T)$ exists for the case of a classical linear spin chain with nearest-neighbor approximation under the  assumption of uniformness, meaning that $J_{i,i+1}$ and $g_i$ are the same for all local spins. 
The solution has the following form\cite{Fisher:class:spin_chain:1964}:
\begin{gather}
\chi(T) = \frac{N (g^{cl})^2 \mu_B^2}{12 kT} \frac{1+u(K)}{1-u(K)}, \\
u(K) = \coth K - (1/K), \\
K = \frac{J^{cl}/2}{kT},
\end{gather}
where $N$ is the number of centers, 
$g^{cl}$ is a classical electron $g$-factor (we assumed $g^{cl} = g^{quant} = 2$), 
$\mu_B$ is Bohr's magneton, and $k$ is the Boltzmann constant.
This expression predicts a maximum on the susceptibility curve at:
\begin{gather}
kT^{\text{cl}}_{max} \approx 0.2382 |J^{cl}|.
\end{gather}

The analytic expression for quantum susceptibility is not known, 
but there are Pad{\'e} approximants\cite{Johnston:pade_approx:2000} fitted to
numerical curves with high quality.  
We used such a curve (Fit 1 in Table I in the Ref. \cite{Johnston:pade_approx:2000})
for calculation of quantum magnetic susceptibility. 
The susceptibility curve of a quantum infinite spin chain has a maximum at
\begin{gather}
kT_{max} \approx 0.640851 |J|. 
\end{gather}
One can immediately see that for the same value of $J^{quant}$, the
classical model yields T$_{max}$, which is about 5.4 times lower that of the quantum model. Conversely,
using the classical model to fit the experimental data would yield a 5.4 times
larger effective
exchange constant. 

It is instructive to compare the predictions of the Heisenberg model with the dimer model,
which gives the well-known Bleaney--Bowers expression for magnetic susceptibility\cite{Bowers:mag_sus:fitting:1952}
\begin{gather}
\chi^{\text{dim}}(T) = \frac{N_A g^2 \mu^2_B}{kT} \frac{2}{3 + e^{-\frac{J}{kT}}}
\end{gather}
The susceptibility of the dimer reaches maximum at
\begin{gather}
kT^{\text{dim}}_{max} \approx 0.624 |J|. 
\end{gather}
Thus, although based on very different physics, the dimer model yields very similar value of T$^{max}$,
i.e., within 3\% from the quantum susceptibility derived from the Heisenberg model. 

\begin{figure}[!h]
\includegraphics[width=7cm]{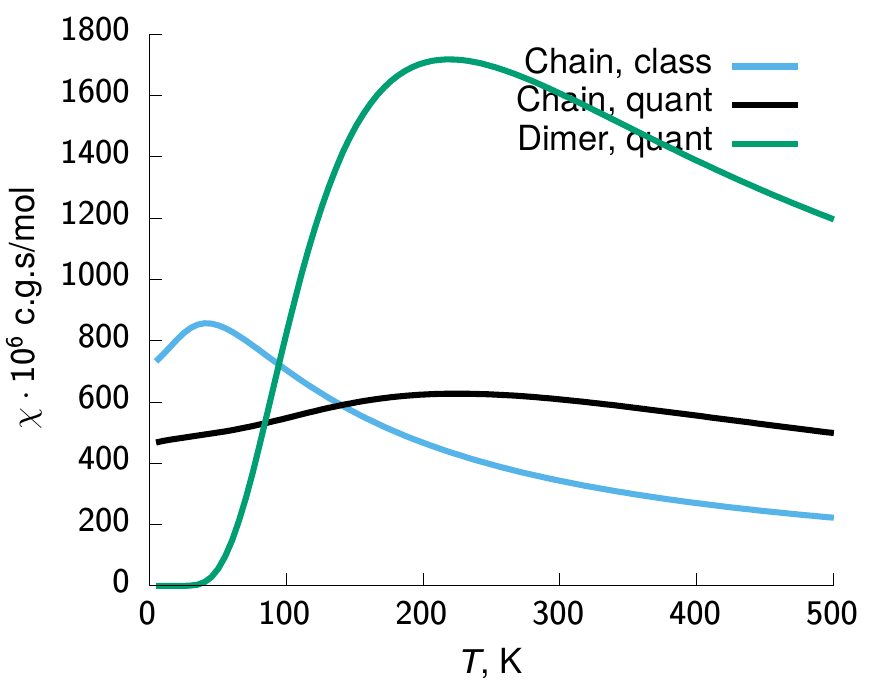} 
\includegraphics[width=7cm]{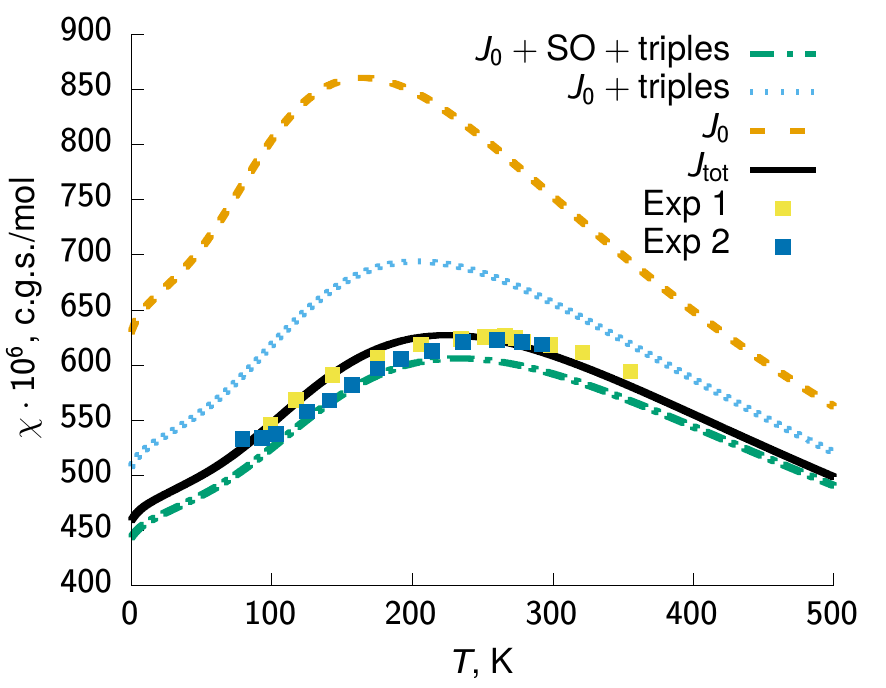} 
\centering
\caption{Left: Classical and quantum susceptibilities of the XXX Heisenberg model and
  the susceptibility for a dimer model (Bleaney--Bowers expression)
  computed with $J=-244$ cm$^{-1}$.
  Right: Quantum susceptibility computed with the effective exchange constants 
 from different approximations. 
         $J_0$ is the converged value of the exchange coupling constant from the 4-center fragment. 
         $J_{\text{tot}}=-244$ cm$^{-1}$ is the overall value of the coupling that includes 
         $J_0$, perturbative triples, spin--orbit shift, and basis-set extrapolation.
         The experimental data set 1 and 2 are from the Refs.~\cite{Dubicki:Cu:chains:1966} and \cite{Zelentsov:Cu:oxalate:1964}, respectively.
  \protect\label{fig:chi_cl_quant}
}
\end{figure}

The results of the classical and quantum Heisenberg models and the dimer model are shown in
Fig. \ref{fig:chi_cl_quant}. 
The susceptibility of the classical spin model differs dramatically from the quantum spin model 
(Figure~\ref{fig:chi_cl_quant}, left), 
making the classical approximation not satisfactory in the interpretation of the magnetic data. 
This is expected, since the susceptibility of the classical model  approaches 
the susceptibility of the quantum model only for the large spin on each center\cite{Xiang:Spin_chains_large_S:1998}.
Classical, quantum, and experimental susceptibility curves 
reach maximum at 42~K, 225~K, and 260~K, respectively. The quantum dimer model gives
the maximum at 219 K.
The difference between the quantum infinite chain and the dimer Heisenberg models in terms of the maximum position of the $\chi(T)$ curve is not
large, but the absolute values are vastly different (the dimer model yields absolute values that are
about 4 times higher).
In the SI (Fig. S4), we show the  comparison between the quantum $\chi(T)$ and the scaled as well as
scaled and shifted dimer's  $\chi(T)$ curves.
As one can see, if the dimer $\chi(T)$ is scaled to match the value of the quantum  $\chi(T)$
at its maximum, then the difference in the slope at low and high temperatures is still clearly visible.
One can scale and shifts dimer's  $\chi(T)$ to match both the maximum and the high-energy part of the curve
(as was done in Ref. \cite{Grivel:CuC2O4:cryst}), but then the slope in the low-energy part remains visibly different.
This analysis clearly shows that the infinite chain model reproduces not only the absolute value of
$\chi(T)$ but also the overall shape of the curve much better than the scaled and shifted dimer model.

We also see that the $\chi(T)$ curve is broad and rather flat, which makes it difficult
to precisely extract $T_{max}$ from noisy  experimental data. 
Hence, the comparison on the basis of the maximum may be misleading, and one should look at the overall
shape. 
Near zero-temperature results would be important to further highlight the discrepancy,
but, unfortunately, there are no reliable data,
as the study that attempted such measurements clearly shows contamination of the sample\cite{Grivel:CuC2O4:cryst}.

The right panel of Figure~\ref{fig:chi_cl_quant} shows the impact of different effects on 
the overall magnetic susceptibility of the quantum spin chain model. 
The value of the effective exchange constant from the 4-center fragment results in 
a susceptibility curve that is far from the experimental one. 
Inclusion of spin--orbit effects and weak correlation through perturbative triples 
improves agreement.
The impact of the basis set is relatively small but it improves agreement even further. 
The more effects that are included, the closer the agreement with the experiment becomes.

Inelastic neutron scattering allows one to probe the manifold of spin states of a magnetic
system. 
The intensity of inelastic neutron scattering is proportional to the dynamic structure factor, given by
\begin{gather}
S^{\sigma\sigma^\prime}(k, \omega) = 
\sum_i \braket{\Psi_{GS} | S_k^{\sigma} | \Psi_i} \braket{\Psi_i | S_k^{\sigma^\prime} | \Psi_{GS}}| 
\delta(\omega - \omega_i),
\end{gather}
where $S^{\sigma}_k = \frac{1}{\sqrt{N}} \sum_j^N e^{iqj} S_j^\sigma$ are 
the spin operators in the periodic representation, 
$\Psi_{GS}$ and $\Psi_i$ are the ground and excited states, 
and $\omega_i$ is the excitation energy of excited state $\Psi_i$. 
Mixtures of weakly interacting antiferromagnetic dimers would  have a sharp intensity peak 
at the transfer energy matching the singlet--triplet gap\cite{Furrer:INS:dimer:1977,Zheludev:dimer:INS:1996}. 
The energy position of the peak does not depend on the momentum transfer $k$.
In contrast, extended spin chains show a characteristic dispersion law, 
where the intensity is large 
at the distribution of energies and wave-vectors\cite{Caux:spin_chain:dyn_str_fact:2005,Kohno:INS:chain:2009}. 
Thus, future neutron scattering experiments for copper oxalate could further
confirm the infinite spin-chain model and tighten the experimental error bar on the effective
exchange constant. 

The rigorously derived {\em ab initio} effective Hamiltonian for copper oxalate corresponds 
to the infinite Heisenberg spin chain model and validates the nearest-neighbor approximation.
This is the first purely theoretical validation
of the XXX Heisenberg model that is based on non-empirical all-electron many-body
calculations and not relying on spin-symmetry broken solutions. 
Our results show that both strong and weak correlation contribute to the effective exchange constant, whereas the dependence on the basis set is relatively weak. 
The spin--orbit interaction introduces a noticeable  additional shift that further changes
the effective exchange constant. 
Only after accounting for all these effects, does
the magnetic susceptibility of the quantum spin model agree well with the
experimental values\cite{Zelentsov:Cu:oxalate:1964,Dubicki:Cu:chains:1966}.
This is the first application of the EOM-SF-CC method to describe
a periodic system, opening a new route in the treatment of 
strongly
correlated periodic systems.
These results provide a solid basis for future quantitative modeling of spin chains, including magnetic field effects. 

\section{Methods}
Cartesian coordinates were extracted from the crystal unit cell parameters of copper oxalate (CuC$_2$O$_4$) 
reported in Ref. \citenum{Grivel:CuC2O4:cryst}.  
The crystal structure is composed of the polymer CuC$_2$O$_4$ (Figure~\ref{fig:all_geometries}). 
The unit cell was repeated and the CuC$_2$O$_4$ stripes were extracted. 
Hydrogen atoms were added to the oxygen ends of the chain with the corresponding to oxalic acid bond lengths and angles. The Cartesian coordinates for all structures are given in the SI. 

All calculations were performed with the Q-Chem quantum chemistry package\cite{qchem_2014,qchem_feature}.
We used Dunning's cc-pVDZ and cc-pVTZ basis sets for all atoms\cite{Dunning:ccpvxz:1989,Duninng:ccpvxz:Sc-Zn,Dunning:ccpvxz:Ca,Dunning:ccpvxz:Ga-Kr}.
Unrestricted orbitals were used in calculations with open-shell references. 
To accelerate electronic structure calculations, we used single precision\cite{Pavel:SP:2018} for CCSD, intermediates, and EOM-SF-CCSD, combined with the \textit{libxm} library for tensor contractions\cite{Kaliman:LibXM:16}. 

We used the open-shell frozen natural orbital truncation of the virtual space\cite{Pavel:OSFNO:2019}
with the truncation threshold 
corresponding to 99\% of the total preserved occupation in the virtual space. 
Core electrons were frozen in all calculations. 
Triples corrections were computed using (fT) perturbative triples\cite{sd3paper} for EOM-SF-CCSD. 
In the  (fT) calculations for a model dimer, the
40 lowest occupied orbitals were frozen for the dimer fragment.

Calculations of spin--orbit couplings\cite{Pokhilko:SOC:19} 
were performed using  EOM-DIP-CCSD wave functions\cite{Wang:DIP:SOC:2020}.

Bloch's and des Cloizeaux's effective Hamiltonians were built for the open-shell model spaces of 
the determinants expressed in the open-shell localized orbitals 
following the procedure from Ref. \cite{Pokhilko:EffH:2020}. 
Localized orbitals were determined using the Foster--Boys criterion of localization\cite{Boys:LMO:60}. 

\section*{Acknowledgments}
We are grateful to Professor Devin Matthews from the South Methodist University for his generous
help with testing the effects of correlation beyond perturbative triples correction. 
This work is supported by the Department of Energy through the 
DE-SC0018910 grant.\\

P.P. current address is: Department of Chemistry, University of Michigan, Ann Arbor, Michigan 48109, USA \\

The authors declare the following competing financial interest(s): A.I.K. is a member of the Board
of Directors and a part-owner of Q-Chem, Inc.

\clearpage
\bibliographystyle{prf}

\end{document}

\title{Is solid copper oxalate a spin chain or a mixture of entangled spin pairs?\\Supplemental Information.}

\author{Pavel Pokhilko$^a$, Dmitry S. Bezrukov$^{b,c}$, and Anna I. Krylov$^a$\\
{\small $^a$ Department of Chemistry, University of Southern California, 
  Los Angeles,  California 90089-0482}\\
{\small $^b$ Department of Chemistry, M.V. Lomonosov Moscow State University, Moscow 119991, Russia}\\
{\small $^c$ Skolkovo Institute of Science and Technology, Skolkovo Innovation Center, Nobel str. 3, Moscow 143026, Russia}\\
}

\maketitle
\section{Effective Hamiltonians} 
\begin{figure}[!h]
\includegraphics[width=6cm]{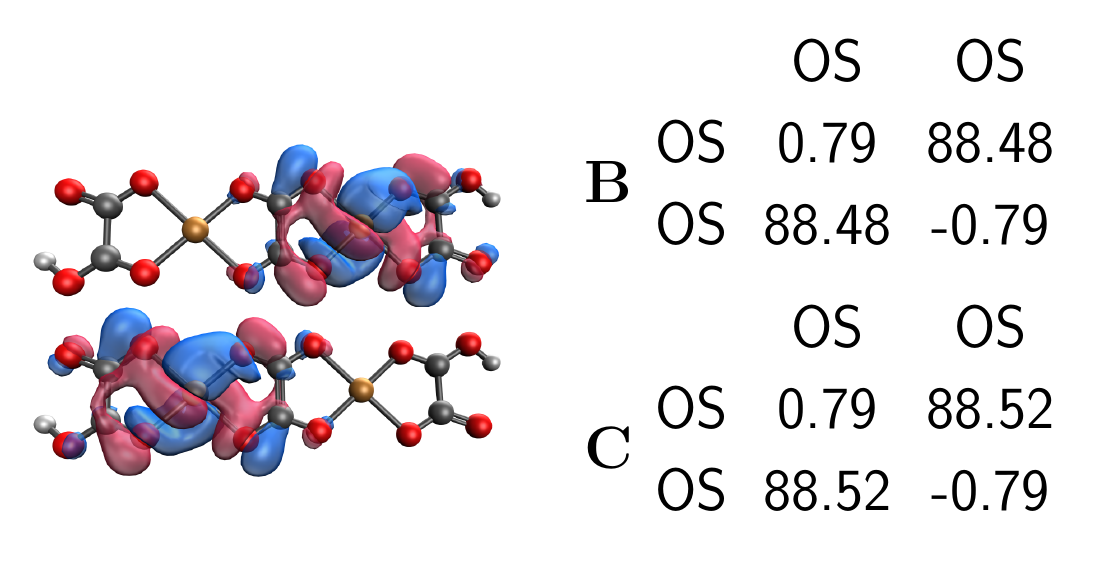} 
\includegraphics[width=8cm]{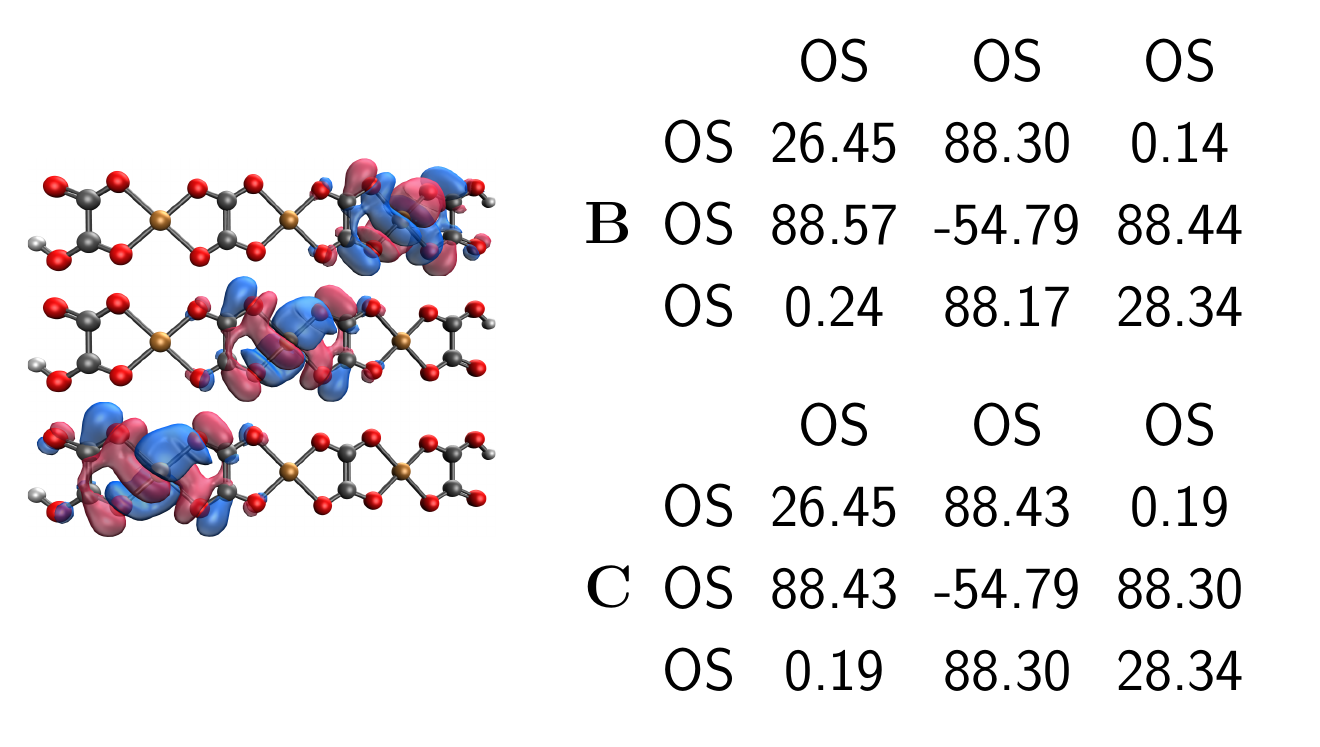} 
\includegraphics[width=9cm]{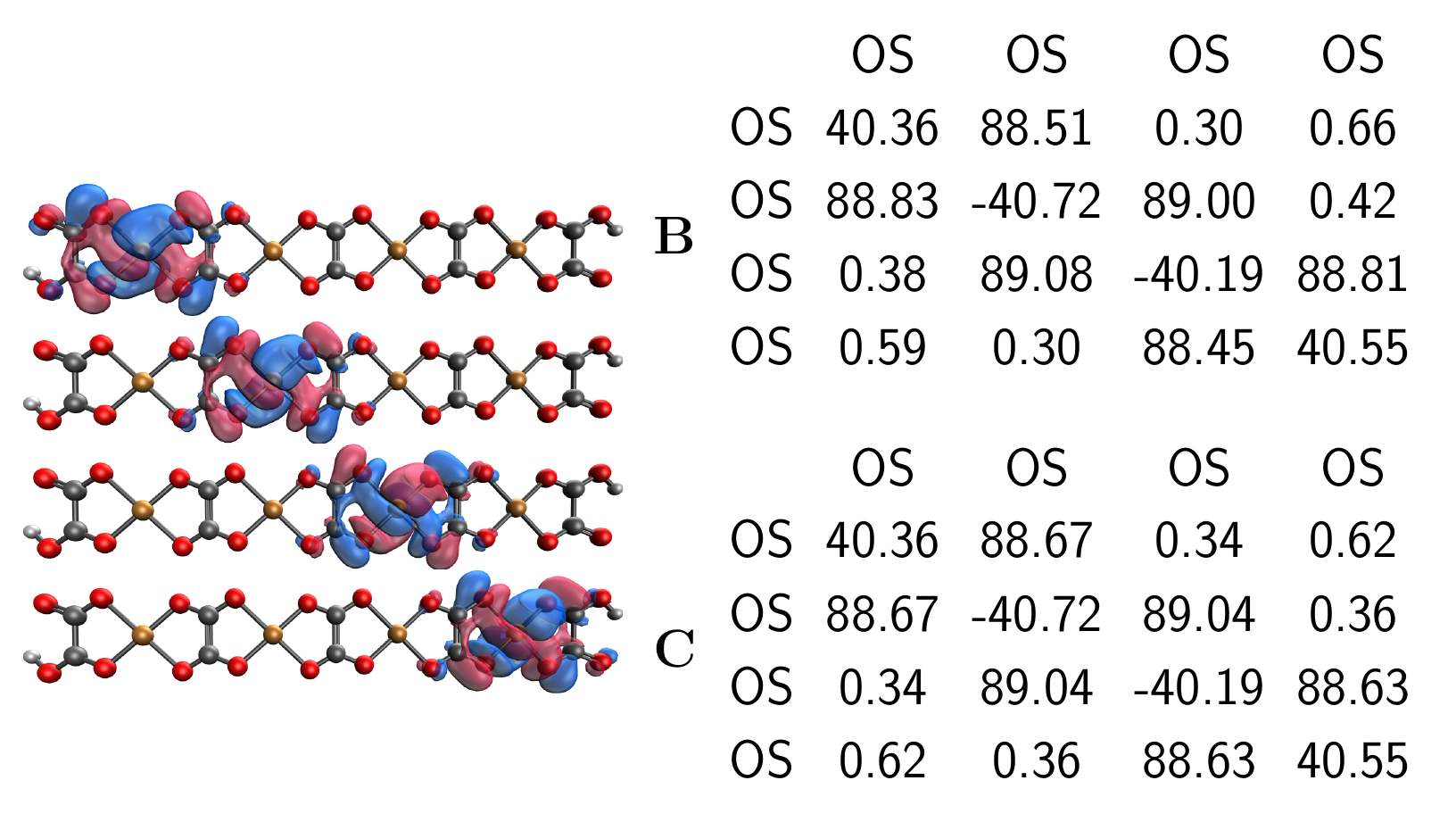} 
\centering
\caption{Bloch's (\textbf{B}) and des Cloizeaux's (\textbf{C}) effective Hamiltonians 
for the bi- and trinuclear fragment. 
         The energies are shifted to produce a zero trace. 
  \protect\label{fig:cu23_eff_ham}
}
\end{figure}

\begin{table}[tbh!]
  \caption{$\left< S^2 \right>$ of the reference determinants used in calculations  
and a comparison with spin-pure solutions. 
\protect\label{tbl:s2_ref}} 
\begin{tabular}{l|c|c|c}
\hline
\# of Cu  & 2     &  3  & 4  \\       
\hline
Spin-pure solution & 2.00  &  3.75  & 6.00 \\
\hline
UHF     & 2.007 &  3.760 & 6.013 \\ 
\hline
\end{tabular}
\end{table}

\begin{table}[tbh!]
  \caption{Effective exchange couplings $J$ extracted with EOM-SF-CCSD/cc-pVDZ. 
Bloch's and des Cloizeax's effective Hamiltonians provide nearly identical values of $J$. 
\protect\label{tbl:J}} 
\begin{tabular}{l|c|c}
\hline
\# of Cu  & Bloch     & Des Cloizeaux  \\       
\hline
2         & -176.96    & -177.04 \\
3         & -176.87, -176.61, -0.38     &  -176.86, -176.60, -0.38 \\
4         & -177.34, -178.08, -177.26, -0.68, -0.72, -1.25 & -177.34, -178.08, -177.26, -0.68, -0.72, -1.24 \\
\hline
\end{tabular}
\end{table}

\begin{table}[tbh!]
  \caption{Impact of truncation to the nearest neighbors on the eigenvalues of the des Cloizeaux effective Hamiltonian. 
Eigenvalues (in cm$^{-1}$) for the full and the truncated Hamiltonians are shown.
\protect\label{tbl:trunc_evals}} 
\begin{tabular}{l|c|c}
\hline
\# of Cu  &  Full & Nearest-neighbor truncation  \\       
\hline
3         & -145.18, 27.20, 117.98 & -145.25, 27.39, 117.86   \\
4         & -167.16, -44.25, 77.50, 133.91 & -167.19, -44.40, 78.15, 133.44 \\
\hline
\end{tabular}
\end{table}

\begin{figure}[!h]
\includegraphics[width=9cm]{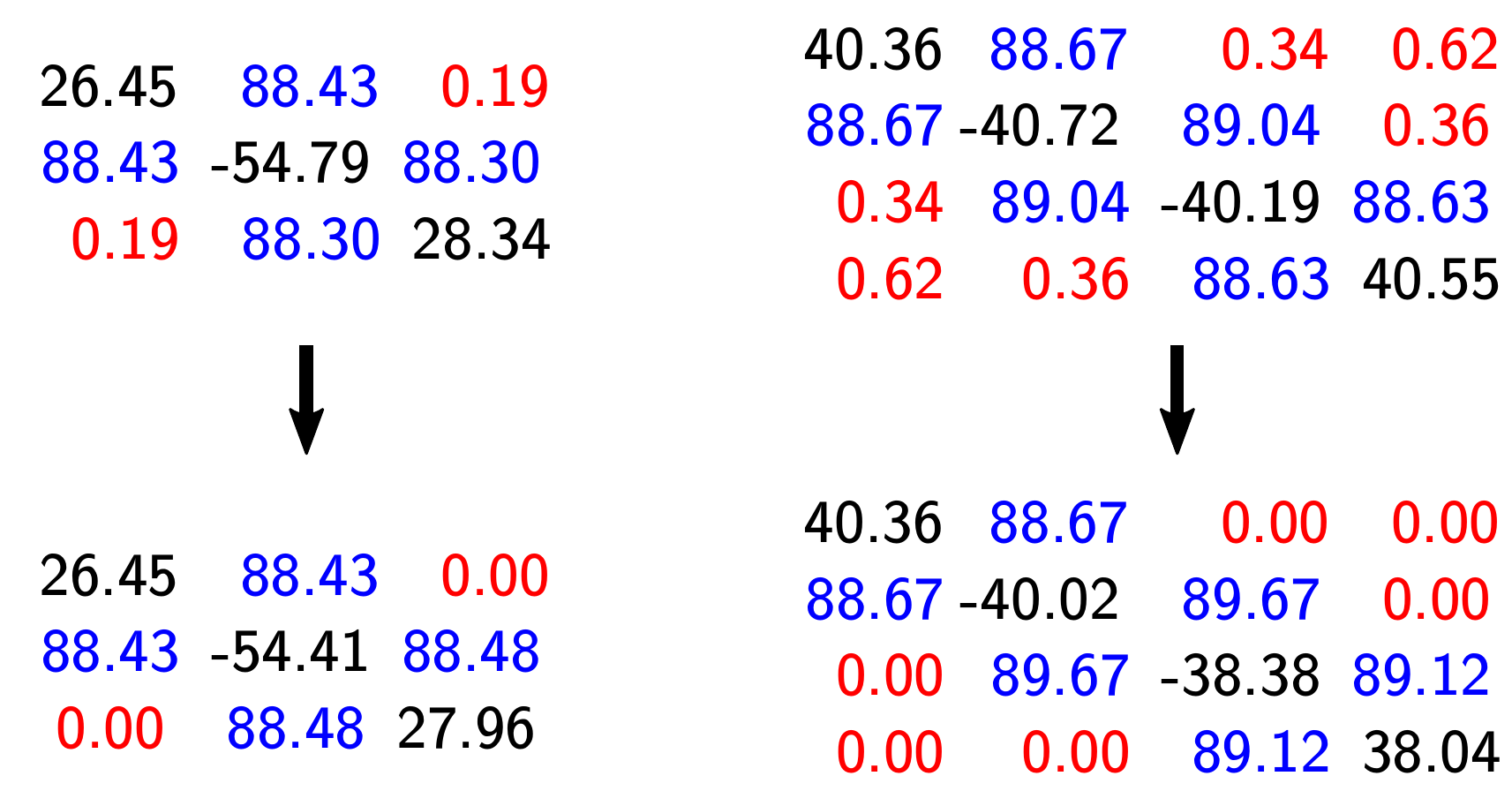} 
\centering
\caption{Tridiagonalization of the des Cloizeaux (\textbf{C}) effective Hamiltonians 
for the tri- and tetranuclear fragment. 
  \protect\label{fig:tridiag}
}
\end{figure}

\clearpage
\section{Spin--orbit calculations}
As we noted in the main text, 
EOM-DIP-CCSD suffers from insufficient treatment of weak correlation and orbital relaxation.
However, the energy shift due to spin--orbit interaction does not depend much on the value of the 
triplet--singlet gap (we report the results with the EOM-SF-CCSD gap in the Table~\ref{tbl:soc}).
This can be rationalized from El-Sayed's rule: 
the Heisenberg states have nearly the same orbital occupancies;
therefore, there is no contribution from the direct coupling between them. 
The relative energy gaps with the other excited states are much larger, 
which makes the spin--orbit shifts insensitive to perturbations in the triplet--singlet gap of hundreds of cm$^{-1}$.

\begin{table}[tbh!]
  \caption{Positions (cm$^{-1}$) of the lowest singlet and triplet states after diagonalization of
    the spin--orbit Hamiltonian of the binuclear fragment. 
    Three components of the triplet states are shown. The last column shows the effect of the spin--orbit
    couplings on the effective exchange constant. 
The zero energy corresponds to the energy of the singlet state prior incorporation of spin--orbit effects. 
The positions of the triplet states has been shifted by the same value prior diagonalization of the spin--orbit Hamiltonians to reproduce EOM-SF-CCSD triplet--singlet gap (177 cm$^{-1}$).
\protect\label{tbl:soc}} 
\begin{tabular}{l|cccc|c}
\hline
\# of states  & S        & T     & T     & T             & $\Delta J$\\       
2S,2T                 & -65.3 & 177   & 177   &  177   &   -65.3   \\   
3S,3T                 & -65.4 & 109.6 & 109.6 &  177   &   -20.5   \\
4S,4T                 & -75.0 & 109.5 & 109.6 & 177    &   -30.0    \\
5S,5T                 & -75.0 & 98.7 & 109.6 & 165.0   &   -22.4   \\
6S,6T                 & -75.1 & 95.1 & 100.3 & 152.6    &  -14.1   \\
7S,7T                 & -89.2 & 95.0 & 100.3 & 152.5    &  -28.1   \\
8S,8T                 & -90.1 & 86.7 & 92.1 & 146.1   &   -21.4   \\
9S,9T                 & -100.1 & 85.7 & 91.5 & 145.5   &   -30.7   \\
10S,10T               & -100.9 & 85.7 & 91.5 & 145.5   &   -31.5   \\
\hline
\end{tabular}
\end{table}

\clearpage
\section{Relationship between classical and quantum Heisenberg models}
\begin{gather}
H^{cl} = -\frac{1}{2}\sum_{i<j} J^{cl}_{ij} \mathbf{s}^{cl}_i \cdot \mathbf{s}^{cl}_j
-\frac{1}{2}\sum_i g^{cl}_i\beta \mathbf{H}\cdot \mathbf{s}_i,  \\
\mathbf{s}^{cl} = \frac{\mathbf{S}^{cl}}{S} 
\end{gather}
where the index \textit{cl} denotes a classical quantity. 
The correspondence between quantum quantities and their classical analogies 
is established through the following relations: %\cite{notation}: 
\begin{gather}
J^{quant} S^2 = \frac{1}{2} J^{cl}, \\
g^{quant} S = \frac{1}{2} g^{cl}.
\end{gather}
For the convenience of the reader, 
we use consistent definition of the Heisenberg Hamiltonian and exchange constants. 
If comparison with other texts shall be made, one should compare definitions of the Hamiltonians. 
For example, we use $H = -\sum_{i<j} J^{our, quant}_{ij} S_i S_j = -\frac{1}{2}\sum_{i\neq j}J^{our, quant}_{ij}S_i S_j$. 
Fisher used $H = -\sum_{i\neq j} J^{Fisher, quant}_{ij} S_i S_j$. Therefore, 
$J^{Fisher, quant} = J^{our, quant}/2$.

\section{Comparison with experiment}

\begin{figure}[h!]
\includegraphics[width=9cm]{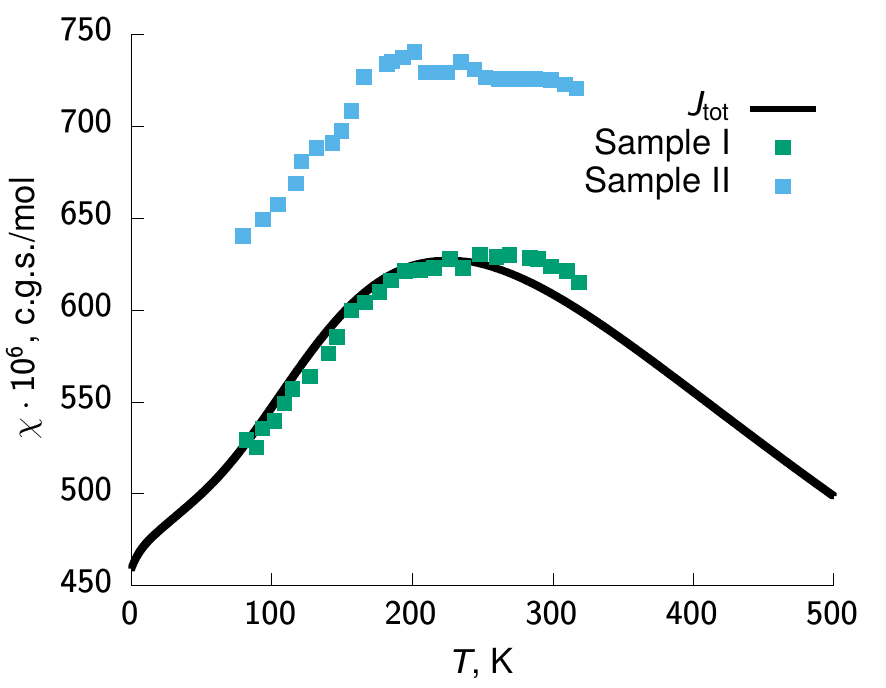}
\centering
\caption{Comparison between our best theoretical estimate ($J_{\text{tot}} = 244$~cm$^{-1}$) 
  and experimental measurements of two samples from the Ref.\citenum{Cu:oxalate:1966}.
  \protect\label{fig:exp_problem}
}
\end{figure}

\begin{figure}[!h]
\includegraphics[width=8cm]{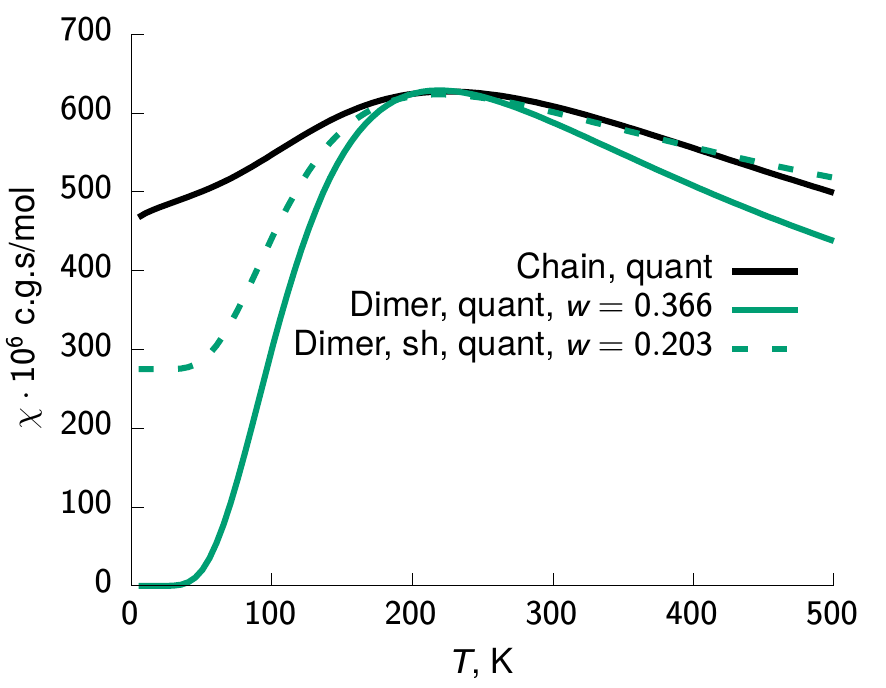}
\centering
\caption{Comparison of susceptibility of the quantum infinite chain and dimer models. 
  Two graphs for the dimer are shown: 
  solid green is the susceptibility of the dimer model that has been scaled by the factor of $0.366$; 
  the dashed line is the susceptibility of the dimer model that has been scaled by the factor $0.201$ 
  and shifted to match the susceptibility of the infinite chain model.
  The weight $0.201$ has been used in the Ref. \citenum{Grivel:CuC2O4:cryst}. 
  The shifted and scaled dimer susceptibility represents the high-temperature region well, but leads 
  to noticeable differences at  lower temperatures.
  \protect\label{fig:scaled_graph}
}
\end{figure}

\clearpage
\renewcommand{\baselinestretch}{1.0} 
\section*{Cartesian coordinates}
\begin{verbatim}
$comment
Fractional atom coordinates within a unit cell.
The last column represents the partial charges, 
used in the work.
$end

Cu   0.00000   0.00000   0.00000   0.35465
 O   0.16200   0.16500   0.80300  -0.22450
 O   0.21500   0.15500   0.39600  -0.14845
 C   0.10900   0.09300   0.55300   0.19560
\end{verbatim}

\section{Relevant Cartesian geometries} 
\begin{verbatim}
$comment
 2-center fragment.
 Nuclear Repulsion Energy =        2461.71150178 hartrees
$end
$molecule                                                                                                                 
0 3
Cu        17.28721       16.65840       13.91353
Cu        15.10708       16.65840       18.55137
C        17.84352       16.14199       11.34880
C        15.66339       16.14199       15.98664
C        13.48326       16.14199       20.62449
C        18.91103       17.17481       11.84041
C        16.73090       17.17481       16.47826
C        14.55077       17.17481       21.11610
O         18.07284       15.74219       10.18934
O         16.86981       15.79772       12.07694
O         15.89271       15.74219       14.82718
O         14.68968       15.79772       16.71479
O         13.71258       15.74219       19.46503
O         12.50955       15.79772       21.35263
O         18.68171       17.57461       12.99987
O         19.88474       17.51908       11.11227
O         16.50159       17.57461       17.63772
O         17.70461       17.51908       15.75011
O         14.32146       17.57461       22.27556
O         15.52448       17.51908       20.38796
H         12.57480       16.30614       22.19545
H         19.76499       17.10589       10.24838
$end
\end{verbatim}

\begin{verbatim}
$comment
 3-center fragment.
 Nuclear Repulsion Energy =        3967.11437581 hartrees                                                                 
$end
$molecule
0 4
Cu        17.28721       16.65840       13.91353
Cu        15.10708       16.65840       18.55137
Cu        12.92695       16.65840       23.18921
O         18.07284       15.74219       10.18934
O         16.86981       15.79772       12.07694
O         15.89271       15.74219       14.82718
O         14.68968       15.79772       16.71479
O         13.71258       15.74219       19.46503
O         12.50955       15.79772       21.35263
O         11.53245       15.74219       24.10287
O         10.32942       15.79772       25.99047
O         18.68171       17.57461       12.99987
O         19.88474       17.51908       11.11227
O         16.50159       17.57461       17.63772
O         17.70461       17.51908       15.75011
O         14.32146       17.57461       22.27556
O         15.52448       17.51908       20.38796
O         12.14133       17.57461       26.91340
O         13.34435       17.51908       25.02580
C        17.84352       16.14199       11.34880
C        15.66339       16.14199       15.98664
C        13.48326       16.14199       20.62449
C        11.30313       16.14199       25.26233
C        18.91103       17.17481       11.84041
C        16.73090       17.17481       16.47826
C        14.55077       17.17481       21.11610
C        12.37064       17.17481       25.75394
H         10.44057       16.21315       26.86015
H         19.76363       17.09192       10.22868
$end
\end{verbatim}

\begin{verbatim}
$comment
 4-center fragment.
 Nuclear Repulsion Energy =        5625.96870548 hartrees                                                                 
$end
$molecule
0 5
Cu        17.28721       16.65840       13.91353
Cu        15.10708       16.65840       18.55137
Cu        12.92695       16.65840       23.18921
Cu        10.74682       16.65840       27.82706
O         18.07284       15.74219       10.18934
O         16.86981       15.79772       12.07694
O         15.89271       15.74219       14.82718
O         14.68968       15.79772       16.71479
O         13.71258       15.74219       19.46503
O         12.50955       15.79772       21.35263
O         11.53245       15.74219       24.10287
O         10.32942       15.79772       25.99047
O          9.35232       15.74219       28.74071
O          8.14929       15.79772       30.62831
O         18.68171       17.57461       12.99987
O         19.88474       17.51908       11.11227
O         16.50159       17.57461       17.63772
O         17.70461       17.51908       15.75011
O         14.32146       17.57461       22.27556
O         15.52448       17.51908       20.38796
O         12.14133       17.57461       26.91340
O         13.34435       17.51908       25.02580
O          9.96120       17.57461       31.55124
O         11.16422       17.51908       29.66364
C        17.84352       16.14199       11.34880
C        15.66339       16.14199       15.98664
C        13.48326       16.14199       20.62449
C        11.30313       16.14199       25.26233
C         9.12300       16.14199       29.90017
C        18.91103       17.17481       11.84041
C        16.73090       17.17481       16.47826
C        14.55077       17.17481       21.11610
C        12.37064       17.17481       25.75394
C        10.19051       17.17481       30.39178
H          8.22335       16.31273       31.45165
H         19.74448       17.10907       10.23780
$end
\end{verbatim}

\clearpage
\bibliographystyle{prf}